\documentclass[twocolumn]{aastex631}  
\usepackage{enumitem}
\usepackage{xcolor}
\usepackage{tablefootnote}
\usepackage{bm}
\usepackage{xcolor}
\usepackage{soul}
\usepackage{placeins}
\usepackage{appendix}
\hypersetup{backref=true, pagebackref=true,  hyperindex=true,colorlinks=blue,breaklinks=true,  urlcolor=violet,linkcolor= red, bookmarks=true, 	bookmarksopen=true,  filecolor=cyan, citecolor=blue, linkbordercolor=blue}
\begin{document}
\title{Investigating the 1L2S Degeneracy for Wide-Orbit Planets}

\author[0009-0007-4190-1269]{PARISA SANGTARASH}
\affiliation{Department of Physics, Isfahan University of Technology, Isfahan 84156-83111, Iran}

\author[0000-0001-9481-7123]{JENNIFER C.YEE$^{\star}$}
\affiliation{Harvard-Smithsonian Center for Astrophysics, 60 Garden street. Cambridge.MA 02138. USA}

\thanks{\textcolor{blue}{\rm{e}-\rm{mail: jyee@cfa.harvard.edu}}}
\begin{abstract}

\noindent
Wide-orbit planets are particularly sensitive to detection by the Roman Galactic Bulge Time Domain Survey (GBTDS). This study investigates the degeneracy of these events with binary sources, focusing on how observation cadence affects the resolution of these degeneracies. We analyzed the impact of various cadences from $(3.6~\mathrm{min})^{-1}$ to $(5~\mathrm{hr})^{-1}$, which encompasses planned cadences for both the GBTDS and cadences used by ground-based surveys like KMTNet. The results show that a $\sim (15~\mathrm{min})^{-1}$ cadence is generally sufficient to resolve this degeneracy unless the source star is a giant ($\rho \gtrsim 0.01$).
\end{abstract}


\section{Introduction}

Relatively few planets on orbits of more than $\sim5$ au are known, especially planets with masses $<1~M_{\text{Jup}}$ (NASA Exoplanet Archive accessed 2025 February 8).
Gravitational microlensing offers a promising method for detecting such planets, as it is less limited by the challenges faced by the transit or radial velocity methods, which are biased toward more massive planets on shorter period orbits.  

In gravitational microlensing events caused by wide-orbit planets, two distinct caustic structures are formed: a \textit{central caustic}, located at the host lens, and a \textit{planetary caustic}, positioned at a distance of $s - \frac{1}{s}$ from the host star. Here, $s$ represents the normalized separation between the host star and the planet, scaled by the Einstein radius, and projected onto the plane of the sky as seen by an observer looking toward the source star. The size and shape of these caustics depend on the host mass ratio ($q$) and their separation ($s$).
When the source star passes through the planetary caustic region, the light curve exhibits characteristic perturbations in the form of bumps or dips, which signal the planet's presence. 

  \citet{poleski2021wide} measured the occurrence rate of wide-orbit planets based on two decades of OGLE data. 
They found that wide-orbit planets, those with $s > 2$ (corresponding to semi-major axes $\sim 5$–$15$ au), 
are relatively common, with an estimated $1.04$ giant planets per Sun-like star, 
with uncertainties of $+0.9$ and $-0.6$. 
However, such planets are still relatively rare, and only one with $q < 10^{-4}$ has been published \citep{2024AJ....167..162Z}.

Wide-orbit planets are of particular interest in the context of the Roman Galactic Bulge Time Domain Survey (RGBS) 
because they represent a distinct parameter space. These events are characterized 
by low-magnification anomalies in the light curves, typically where the magnification is $A \approx 1$. 
RGBS is more effective at detecting such low-amplitude features compared to ground-based surveys like KMTNet
\citep{gould2023kmt}. 

  However, low-amplitude perturbations from binary sources can mimic those caused by wide-orbit planets. Therefore, these models can be degenerate models for wide-orbit planetary events \citep{gaudi1998distinguishing}.
  Thus, the study of degeneracies associated with wide-orbit planetary microlensing events is essential.  \citet{yee2023scientific} discuss potential issues regarding the degeneracy of wide-orbit planets with other models in the context of RGBS.
In this paper, we aim to explore the binary-source degeneracy as a step toward resolving these challenges.  
To this end, we have investigated the crucial role of cadence in reducing or exacerbating these degeneracies.

  Here, we focus on how the observational cadence and various parameters of the microlensing event affect
our ability to resolve this type of degeneracy, testing a variety of cadences allows us to compare Roman’s ability to  
resolve these degeneracies relative to current surveys with ground-based  
telescopes.

Section 2 describes the method for simulating wide-orbit planetary events.  
Section 3 explains the procedure for fitting binary sources to simulated wide-orbit planetary events.  
Section 4 presents the results regarding the degeneracy between binary sources and wide-orbit planets under different parameters and conditions.  
In Section 5, we discuss the potential for resolving the degeneracy based on the flux ratio of the fitted binary sources.  
In Section 6, we show how our results can be scaled for different assumptions.
Section 7 is dedicated to the conclusion.  

\section{SIMULATED DATA}

\begin{figure*}[htbp]
\centering
\includegraphics[width=0.44\textwidth]{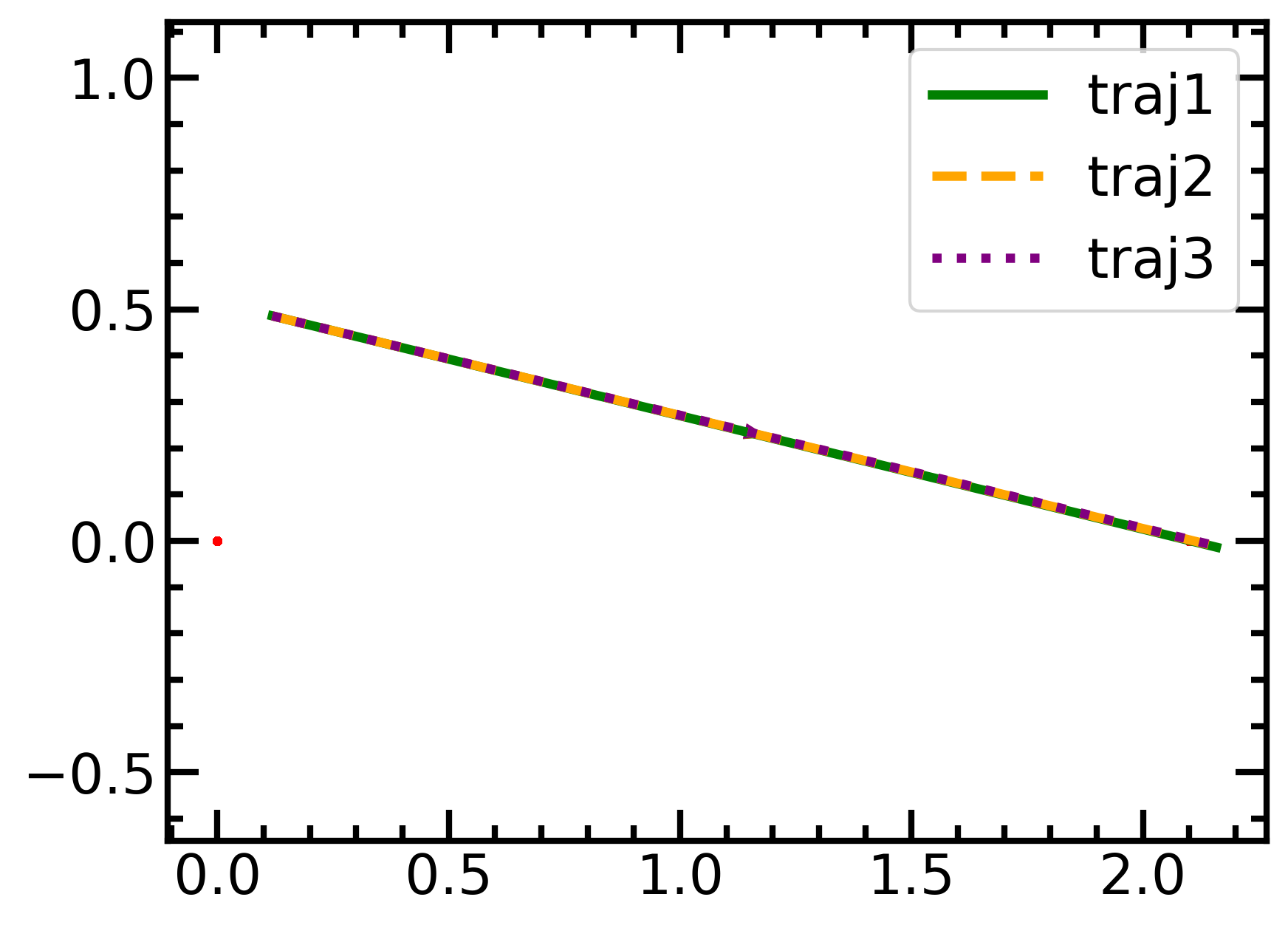}
\includegraphics[width=0.483\textwidth]{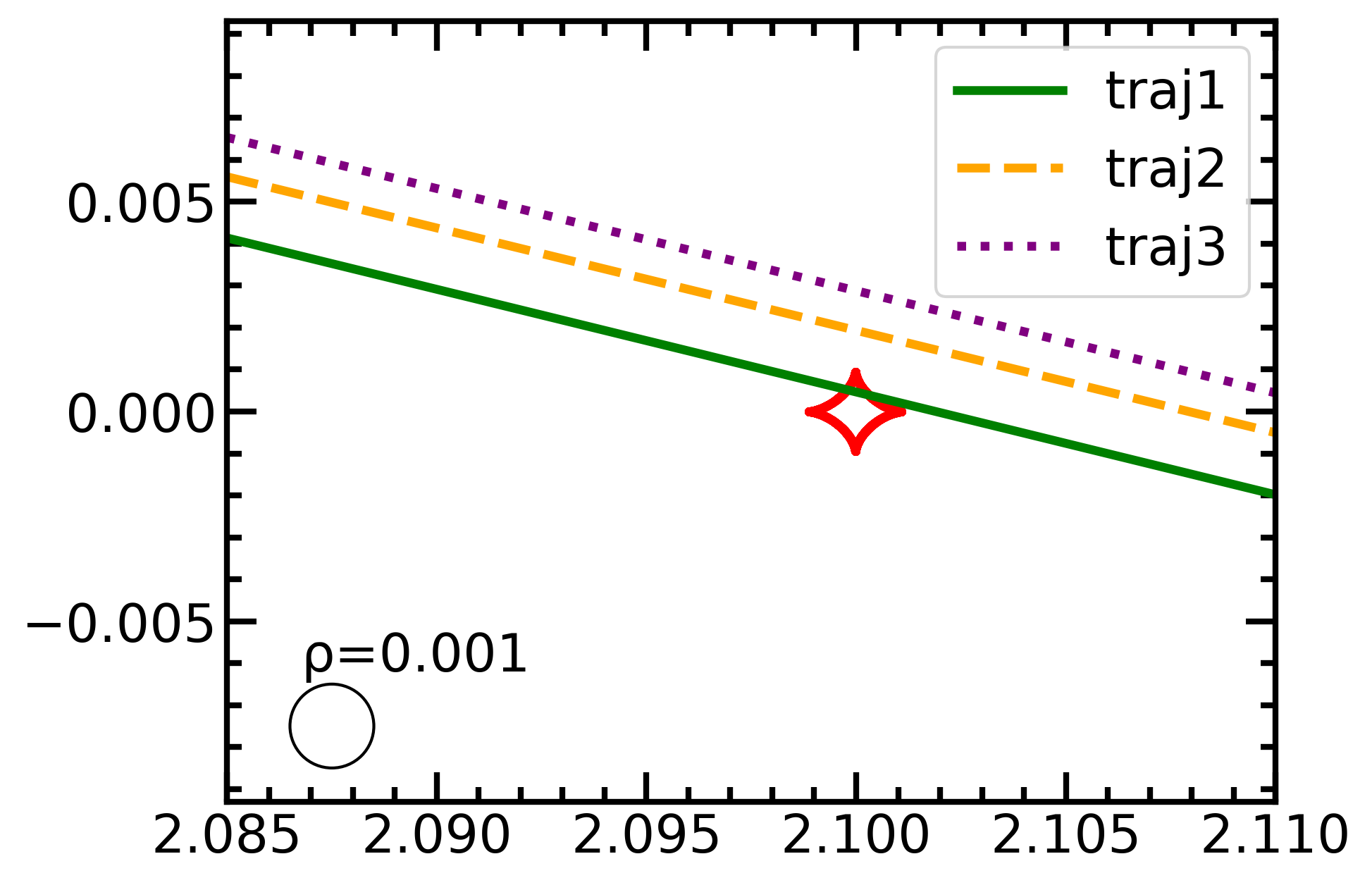}
\caption{Different trajectories of the source with respect to the planetary caustic for three distinct cases. 
\textit{traj1} represents the case where the source passes through the caustic. 
\textit{traj2} corresponds to the case where the source passes tangentially to the upper cusp of the caustic. 
\textit{traj3} depicts the case where the source passes outside but near the upper cusp of the caustic. The right panel is a zoomed-in view of the left panel around the planetary caustic region. In the bottom-left corner of this panel, the size of the source star is shown. In this configuration, the parameters are as follows: 
$s = 2.5$, $\log(q) = -5$, $u_0 = 0.5$, $t_E = 20$, and $\rho = 0.001$.
}
\label{fig1}
\end{figure*}

  To explore the effect of cadence on our ability to resolve the binary-source degeneracy for  
wide-orbit planets, we created simulated light curves covering a variety of microlensing  
event parameters.
  In our simulations, we considered logarithmic mass ratios \( \log(q) = -4, -4.5, -5, -5.5, -6 \), which align with the expected values for planetary systems.
  To ensure that the simulated events fell within the wide-orbit regime, we selected values of \( s > 1.5 \). Specifically, we used \( s = 1.5, 2, 2.5, 3.0, 3.5 \). This choice ensured that the source-lens separation was large enough to produce wide-orbit planetary events \citep{yee2021ogle}.
To analyze how varying \( \rho \) influences the degree of degeneracy with binary sources, we selected values of \( \rho = 0.01, 0.001, \text{ and } 0.0001 \) to systematically investigate the impact of source size on the observed anomalies.

The trajectory angles in our simulations were calculated using the equations from \cite{han2006}, which provide critical information on the longitudinal and transverse dimensions of the planetary caustic, as well as its distance from the host star, as formulated in \cite{bozza1999caustics}.  Using this information, we calculated three angles such that the source could either pass through the caustic, tangentially graze the cusp, or pass outside and near the caustic. 
The measurements provided in \cite{han2006} are approximations for planetary mass ratios, but their accuracy improves as the mass ratio decreases. Given the small values of \( q \) used in our study, these approximations are sufficiently precise.

  \citet{han2006} defines $\Delta \eta$ to be the transverse size of the planetary caustic for wide-orbit events. 
We defined the trajectories for the three scenarios as follows:
\begin{itemize}
    \item \textbf{\textit{traj1}} corresponds to trajectories that cross the caustic at a transverse distance of $\Delta \eta / 4$ relative to the binary axis.
    \item \textbf{\textit{traj2}} corresponds to trajectories that tangentially graze the upper cusp of the caustic with a transverse offset of $\Delta \eta / 2 + \rho$, accounting for the finite size of the source.
    \item \textbf{\textit{traj3}} corresponds to trajectories that pass outside the caustic and near it, with a transverse distance of $\Delta \eta + \rho$.
\end{itemize}
This method ensured that the simulated trajectories accurately represented the desired interactions with the planetary caustic across all three scenarios.
These three scenarios relative to the caustic are illustrated in Figure \ref{fig1} for the case where $s = 2.5$, $\log(q) = -5.0$, $u_0 = 0.5$, and $\rho = 0.001$.

  We chose the impact parameter values $u_0 = 0.3, 0.5, 0.7$. These specific values were chosen to ensure that the source trajectory avoids the central caustic.
In these simulations, the primary effect of these different values of $u_0$ is to change the angle of the source trajectory with respect to the planetary caustic, thereby varying the exact form of the planetary perturbation.

  We set $t_0$, the time of the main peak of the light curve, to zero in our simulations.  
For simplicity, and to enable simple scalings of our results to other conditions,  
we use a timescale of $t_E = 20$ days, which is representative of the typical event duration in the Roman Telescope’s survey field.

\begin{figure*}[htbp]
\centering
\includegraphics[width=1.0\textwidth]{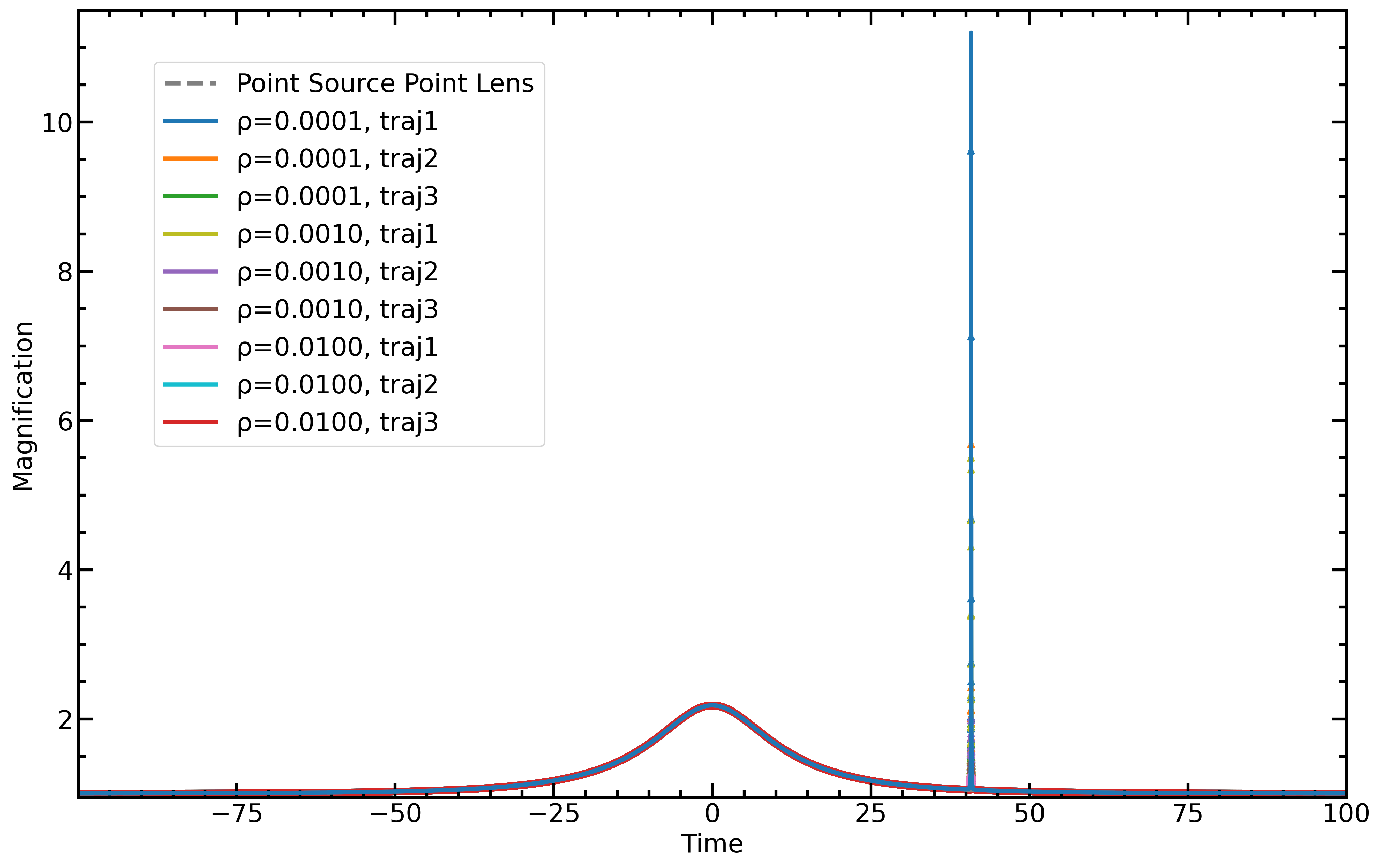}
\includegraphics[width=0.49\textwidth]{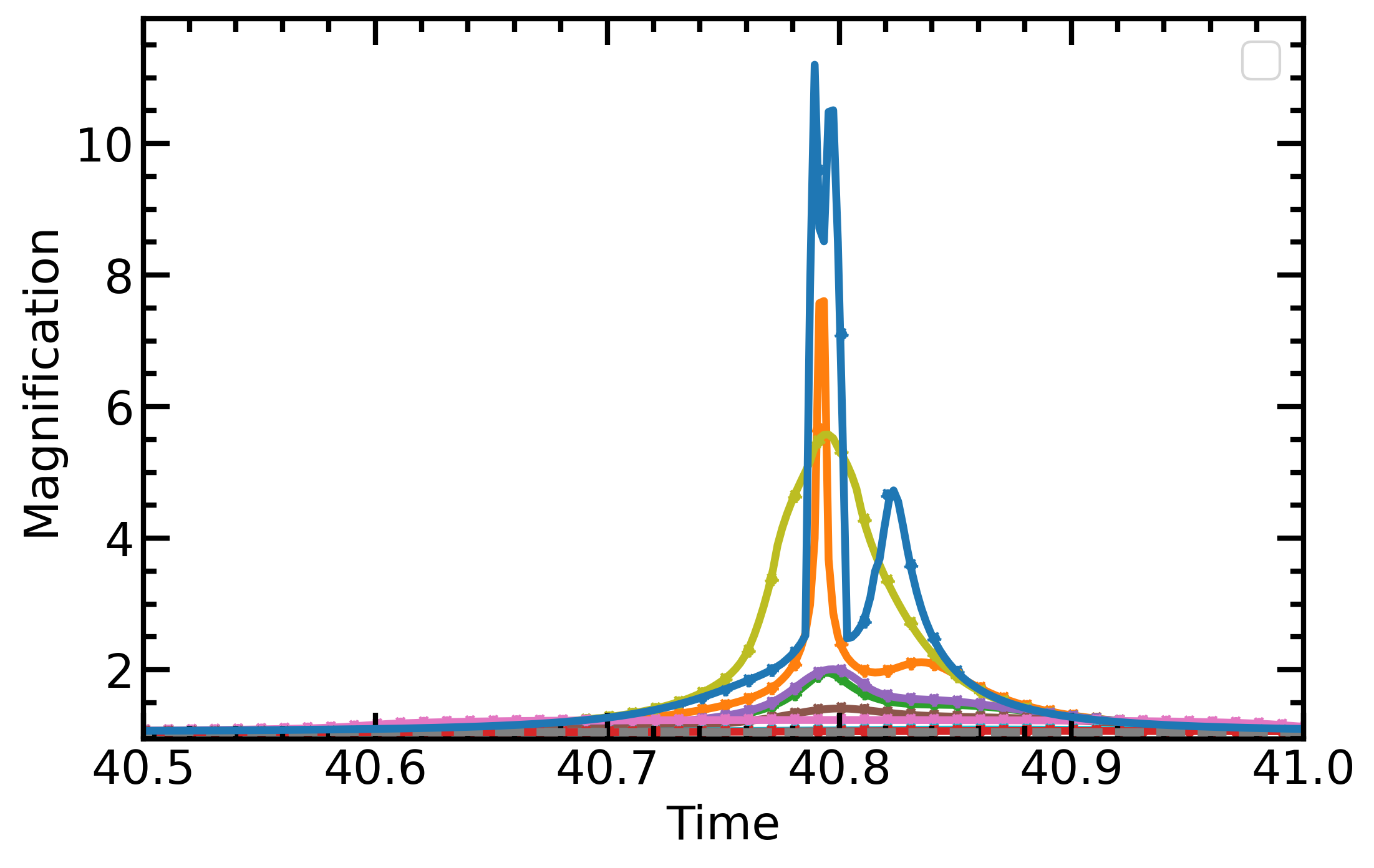}
\includegraphics[width=0.49\textwidth]{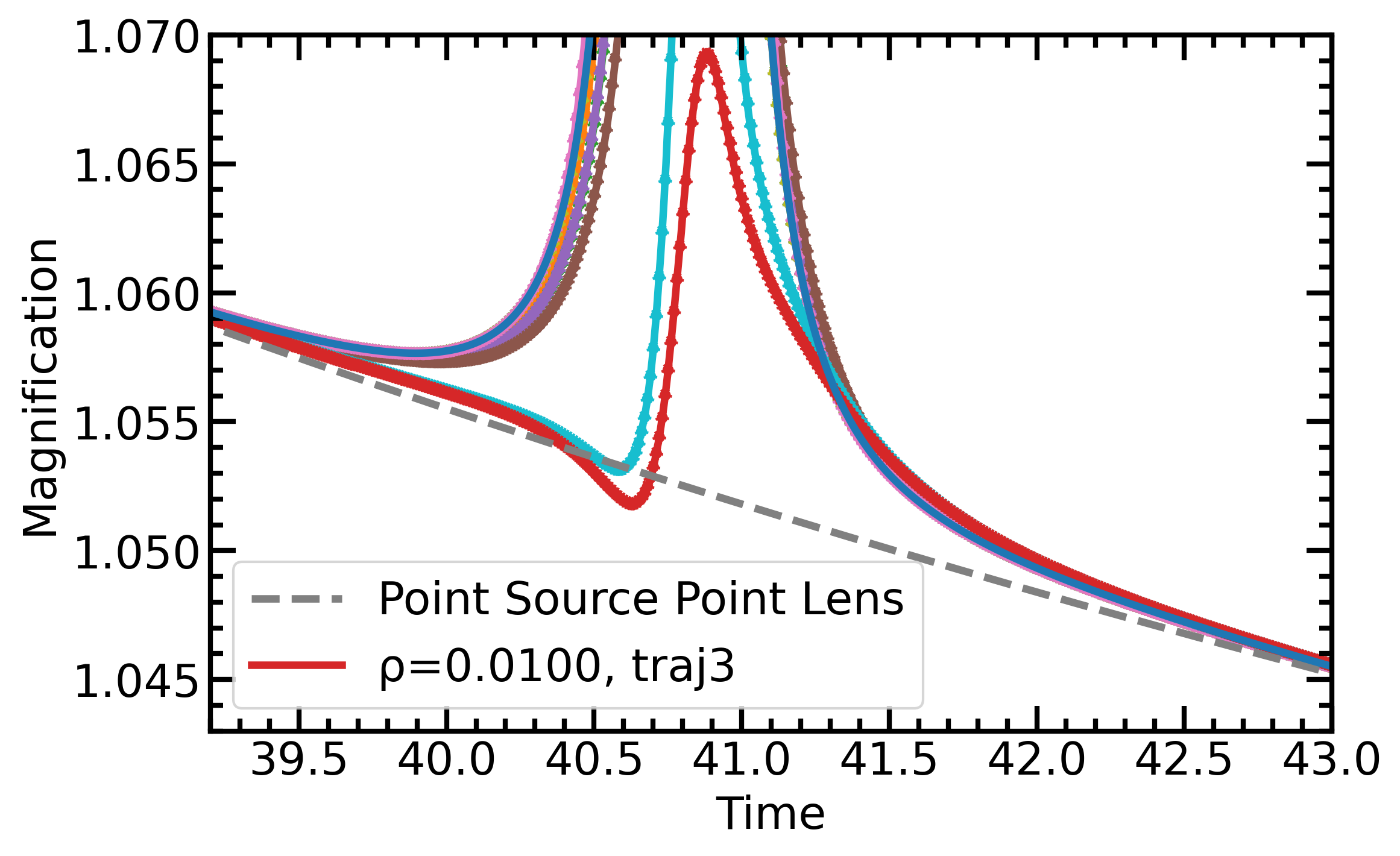}
\caption{Light curves for the parameters 
$t_0 = 0$, $u_0 = 0.5$, $s = 2.5$, and $\log(q) = -5$, 
calculated for three different values of $\rho = 0.01$, $0.001$, and $0.0001$, and for three different trajectories \textit{traj 1}, \textit{traj 2} and \textit{traj 3}
($3 \times 3 = 9$ cases in total). The top panel shows the full simulated light curve.
The two bottom panels zoom in on the planetary bump-like perturbations. The left panel shows a zoomed view of all 9 bump-like perturbations, while the right panel focuses on the smallest bump. 
In these panels, the data points correspond to a case with 20,000 data points, and these points are overlaid on the different models. Additionally, the point-source point-lens model is represented by a dashed gray line. In these light curves, error bars and fitted binary source models are not displayed.
}
\label{fig3}
\end{figure*}

  In total, the combinations of parameters (five values of $s$, five values of $q$, three values for $\rho$, $\alpha$, and $u_0$) result in 675 total simulated light curve models.  
Figure \ref{fig3} presents light curves for $u_0=0.5$, $q=10^{-5}$, and $s=2.5$, calculated for different trajectories and $\rho$ values.  
The light curves are consistent with one another except at the locations of planetary perturbations.  
The zoomed-in regions in the lower panels highlight these perturbations, showing the differences in bump shapes for the three distinct trajectories and three different $\rho$ values.
In the lower-right panel, the bump for \textit{traj3} with the largest $\rho$ value, representing the smallest bump, is further magnified.

  Within the time interval from $-5t_E$ to $+5t_E$, we sampled each light curve with varying numbers of data points, $N_{pts}$, corresponding to different cadences.

\begin{itemize}
    \item \textbf{1000 data points}: The time interval between consecutive points is $4.8$ hours, which rounds to approximately $5$ hours.
    \item \textbf{2000 data points}: The time interval between consecutive points is $2.4$ hours, which rounds to approximately $2.5$ hours.
    \item \textbf{5000 data points}: The time interval between consecutive points is $0.96$ hours, which rounds to approximately $1$ hour.
    \item \textbf{20,000 data points}: The time interval between consecutive points is $14.4$ minutes, which rounds to approximately $15$ minutes, consistent with Roman's cadence.
    \item \textbf{40,000 data points}: The time interval between consecutive points is $7.2$ minutes.
    \item \textbf{80,000 data points}: The time interval between consecutive points is $3.6$ minutes.
\end{itemize}

Technically, these values are the number of intervals between data points, and the actual number of data points is one large. For example, a cadence of $(0.005~\mathrm{day})^{-1}$ over a total duration of \(200\) days (from \( -5\,t_E \) to \( +5\,t_E \)
), the number of data points can be calculated as \(200 / 0.005 = 40,000\). Including the first data point, we would have \(40,000 + 1 = 40,001\) data points. From now on, for simplicity, we will refer to the total number of data points as \(40,000\), but it should be understood that this actually represents \(40,001\) data points.

  The selected data points were chosen to encompass both a range of possible cadences for the Roman Space Telescope and those of ground-based telescopes like KMTNet  
(cadence = 15 minutes, 1 hour, 2.5 hours, and 5.0 hours).  
This is useful for analyzing the effect of cadence on breaking the degeneracy for a variety of potential real-world situations.

  For our simulated data, we used several fixed parameters (in addition to $t_E$) to simplify the investigation and enable scaling to other situations.  
We used a source flux of 1 and a blend flux of zero.  
We assumed 1\% uncertainties in the measured flux.  
We assumed a uniform source and, therefore, did not include limb-darkening in our calculations.  
The combination of these three assumptions allows our results to be applied to any source star given a particular photometric precision, provided that the limb-darkening effect is not severe.  
We also did not add any noise to the simulated observations.

\section{FITTING BINARY SOURCE MODELS}\label{sec2}

  The lensing of light from two sources by a single lens produces a light curve that includes a primary peak, corresponding to the lensing of the primary source, and a 
second peak
caused by the lensing of the secondary source.
When the flux ratio of the secondary source to the primary ($\varepsilon$) is very small (on the order of $10^{-2}$ to $10^{-4}$),  
meaning the secondary source is significantly fainter than the primary, and the impact parameter of the secondary source ($u_{02}$)  
relative to the lens is very small ($u_{02} \leq \varepsilon / \delta_{\max}$),  
the secondary peak can be similar to a planetary perturbation \cite{gaudi1998distinguishing}.  
In such cases, $\delta_{\max}$, which represents the fractional deviation of the magnification from  
the unperturbed magnification, typically ranges between 5\% and 20\% of the primary peak.

  Given that the detection rate of binary source events  
is expected to be similar to the occurrence rate of planetary events \citep{gaudi1998distinguishing},  
investigating the potential degeneracy with wide-orbit planets is of significant importance.  
To examine this degeneracy, we fit  
the simulated wide-orbit planet light curves with binary source models.

  For a binary source, the flux is simply the superposition of the individual fluxes corresponding to the primary and secondary sources \citep{griest1992effect}. The observed
fluxes for these two sources are \( F_1 = A_1 F_{01} \) and \( F_2 = A_2 F_{02} \), where \( A_1 \) and \( A_2 \) are the magnifications for the primary and secondary sources, respectively. 
  If the flux ratio of the secondary source to the primary source is denoted by 
$\varepsilon = \frac{F_{02}}{F_{01}}$, the total magnification can be written as:
\[
A_{\text{tot}} = \frac{A_1 + \epsilon A_2}{1 + \epsilon}
\]
  In this formulation, the magnifications of the individual sources, $A_1$ and $A_2$,  
are determined by the standard single-lens, single-source equation \citep{1986ApJ...304....1P}.
  The relevant parameters are
\begin{itemize}
   \item $u_{01}$: the impact parameter of the primary source with respect to the lens.  
    \item $t_{01}$: the time of closest approach between the lens and the primary source; qualitatively, it is the time of peak magnification.  
    \item $u_{02}$: the impact parameter of the secondary source with respect to the lens.  
    \item $t_{02}$: the time of closest approach between the lens and the secondary source; qualitatively, it is the time at which the anomaly (or perturbation) occurs in the light curve.  
    \item $t_{\rm E}$: the Einstein timescale for the gravitational microlensing of a binary source by a single lens.
    \item $\rho_2$ is the normalized source size of the secondary source, expressed in units of the Einstein radius. 
\end{itemize}
We have assumed that the sources are stationary.

  We used the \texttt{MulensModel} package \citep{poleski2019modeling} to fit the binary source models to the simulated data.  
Using this package to fit the lensing parameters is straightforward,  
as it first performs a linear regression to fit the blend flux and source flux to the data. 
Because we have two sources, the package fits the flux for both sources.  
We fit for the remaining parameters using \texttt{MCMC (Markov Chain Monte Carlo)}.

  The initial values for the parameters $u_{01}$, $t_E$, and $t_{01}$ for the primary source  
are set to the values from the wide-orbit planetary model used to generate the simulated data.  
We also used the parameters of the wide-orbit planetary model and the equations of \cite{han2006}  
to calculate the peak time of the planetary perturbation, which we used as an initial estimate for $t_{02}$.  
For the secondary source, we chose an initial value of $u_{02} = 0.001$ as a small value,  
and we used the value of $\rho$ from the wide event as the initial value for $\rho_2$.  
The initial parameter values were selected by adding random noise  
from a normal distribution to the starting values. We used  $
\sigma = [0.1, 0.01, 0.01, 0.01, 0.01, 0.001]
$ for the widths of the distributions for $t_{01}$, $u_{01}$, $t_E$, $\log q$, $t_{02}$, and $u_{02}$, respectively.  
Specifically, random values were taken using  
\texttt{Numpy.random.randn(n\_dim)} (\texttt{n\_dim} represents the number of parameters for the fit) and scaled by the $\sigma$ values.  
We chose 20 walkers and set the number of steps to 500 for the fitting process.

\begin{figure*}
	\centering	
	\includegraphics[width=\textwidth]{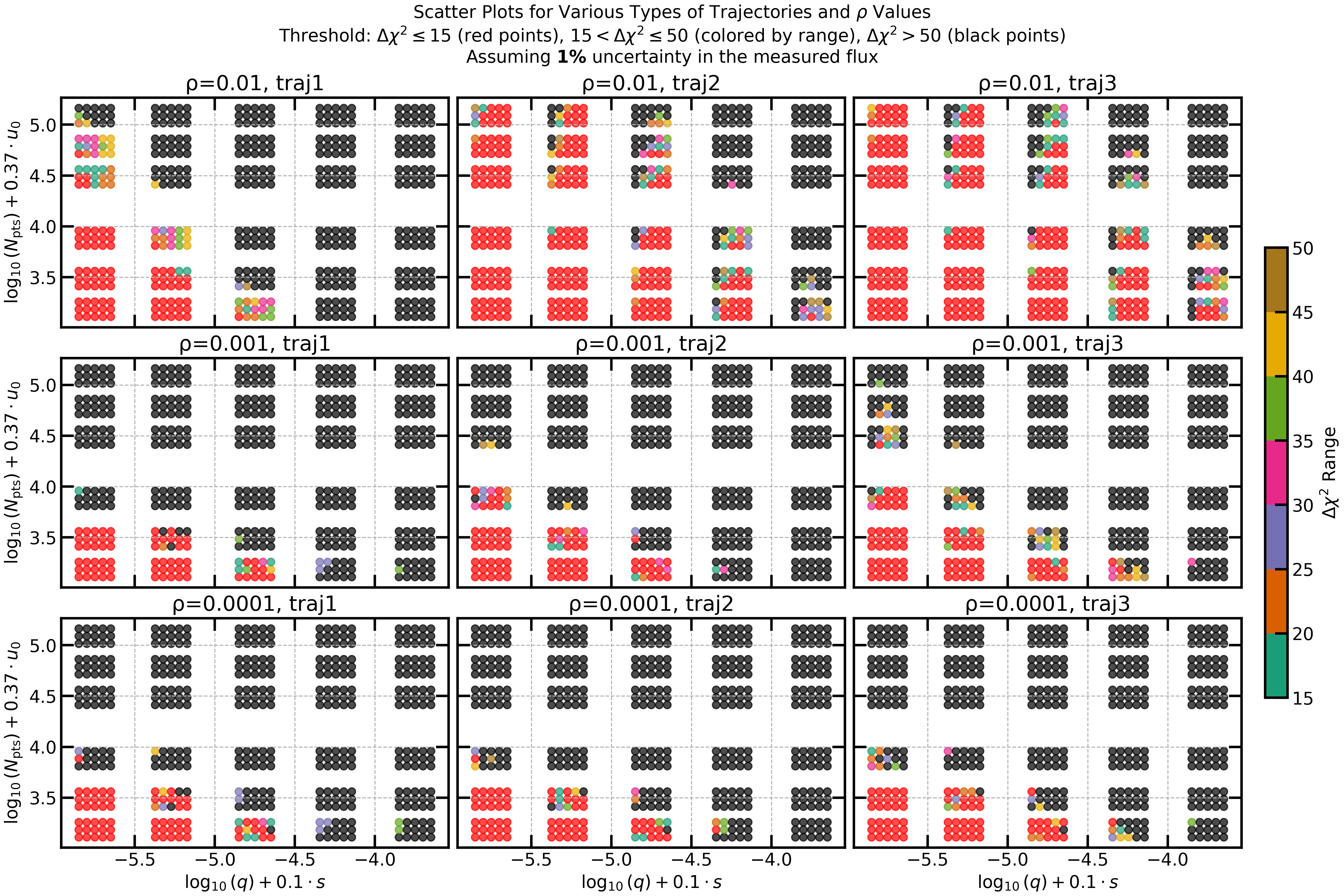}
	\caption{Scatter plots of simulated wide planetary microlensing parameters for different values of $\rho$, $q$, $s$, and $u_0$ across varying trajectories of the source. 
        Each panel (total of 9 panels) corresponds to a specific trajectory(\textit{traj 1}, \textit{traj 2}, and \textit{traj 3}) and $\rho$ value ($\rho$= \{0.01, 0.001, 0.0001\}).
        The horizontal axis represents $\log_{10}(q) + 0.1s$ where \(\log q = \{-6, -5.5, -5, -4.5, -4\}\) and \(s = [1.5, 2, 2.5, 3, 3.5]\) while the vertical axis shows $\log_{10}(\text{Number of Data Points}) + 0.37u_0$ where \( N_{\text{pts}} = \{1000, 2000, 5000, 20000, 40000, 80000\} \) and \(u_0=\{0.3, 0.5, 0.7\}\).
        In each panel, there are 30 groups of points, with each group containing 15 points. These points correspond to combinations of three values of $u_0$ and five values of $s$ ($3 \times 5 = 15$).
        The value of $\Delta \chi^2$ represents a measure of the deviation of the binary source model from the wide-orbit planet model under our assumption of 1\% photometric errors.
    }
	\label{fig1t}
\end{figure*}

\section {RESULTS}\label{sec3}

  The scatter plots in Figure \ref{fig1t} show the results of our simulations and fitting.  
Each of the nine panels corresponds to a specific $\rho$ and a specific trajectory (\textit{traj 1}, \textit{traj 2}, and \textit{traj 3}).  
The x-axis is given by  
$
\log_{10}(q) + 0.1s,
$ 
and the y-axis is  
$
\log_{10}(N_{\rm pts}) + 0.37u_0
$
where the number of data points is denoted by $N_{\rm pts}$.
This groups points such that each $3 \times 5$ block of 15 points corresponds to a specific value of  
$\log q$.  
Within a given $3 \times 5$ block, $s$ increases to the right, and $u_0$ increases upward.  

  Because the simulated data have no noise, the $\chi^2$ of the best-fitting binary source model  
corresponds to the $\Delta\chi^2$ between the wide-orbit and binary-source models.  
We considered $\Delta\chi^2$ values $\leq 15$ to indicate  
an unresolvable degeneracy and $\Delta\chi^2$ values $> 50$ as the  
threshold above which the degeneracy between the binary source model  
and the wide-orbit planetary model is resolved.  
Consequently, we focused on the range of $15 < \Delta\chi^2 \leq 50 $ 
as the region of interest and examined the parameters and trajectories  
for which this degeneracy could be resolved for each cadence.  

  Examining the results for $N_{\rm pts} = 20,000$ data points ,  
which is similar to the cadence expected for the Roman telescope ($15.16$ min), we find:  
\begin{enumerate}
    \item The degeneracy is most severe for larger $\rho$, particularly at $\rho = 0.01$,  
    while it diminishes for smaller $\rho$ and is completely resolved in all cases for $\rho = 0.0001$.  
    \item At $\rho = 0.01$, where degeneracy is present across all trajectories (\textit{traj 1}, \textit{traj 2} and \textit{traj 3}),  
    the fraction of degenerate cases increases for smaller $q$.  
    \item For \textit{traj 1} (caustic crossing trajectories) and $\rho = 0.01$:  
    degeneracy is only present at $\log_{10}(q) = -6$,  
    and the majority of these cases have $\Delta\chi^2 > 15$.  
    \item For \textit{traj 2} (caustic grazing) and \textit{traj 3} (non-caustic crossing),  
    the fraction of degenerate cases increases.  
\end{enumerate}
These general trends apply to all cadences,  
with increasing values of $N_{pts}$ corresponding to decreases in degeneracy and vice versa, as one would expect.  

  Comparing the light curves in Figure \ref{fig3} reveals that bumps associated with  
larger $\rho$ values and trajectories corresponding to \textit{traj 2},  
and especially \textit{traj 3}, are shorter in height and exhibit  
smoother, simpler shapes. Therefore, we expect that these cases  
are more likely to be subject to the binary source degeneracy.  
We find that this is true in practice:  
events associated with \textit{traj 1}, characterized  
by trajectories passing through the caustic, exhibit significantly  
lower rates of degeneracy compared to events  
associated with \textit{traj 3}, which involve non-caustic trajectories.  
Additionally, the degeneracy intensity consistently  
increases for smaller values of $q$ and larger values of $\rho$.

  In another analysis of the results, the minimum number of data points  
(or the minimum cadence) required for degeneracy to be resolved  
can be investigated for different values of $q$.  
Based on the observations in Figure 2, for events  
with $\log_{10}(q) = -4$, the minimum number of data points 
required to resolve the degeneracy between binary sources  
and wide-orbit planets is:  
\begin{itemize}
    \item For \textit{traj 1} events: 1,000  
    \item For \textit{traj 2} events: 2,000  
    \item For \textit{traj 3} events: 5,000  
\end{itemize}
In relation to $\log_{10}(q) = -4.5$:  
\begin{itemize}
    \item For \textit{traj 1} events: 2,000  
    \item For \textit{traj 2} events: 20,000  
    \item For \textit{traj 3} events: 40,000  
\end{itemize}

\section{Physical Constraints from Flux Ratio}

From the light curve, the two measured source properties are the source flux and $\rho$. Combining a measurement of $\rho$ with the intrinsic size of the associated source, $\theta_*$, inferred from the flux, yields the value of the lens-source relative proper motion, $\mu_{\rm rel} = \theta_{\rm E} / t_{\rm E}$ where $\theta_{\rm E} = \theta_* / \rho$. When the source is similar in size to or larger than the caustic structure, the duration of the perturbation is set by the source size. Hence, the measured value of $\rho$ for the best-fitting binary source model should be similar to the true value for the planetary light curve. 

However, the binary source degeneracy requires the binary source to have an extreme flux ratio. In this scenario, the star producing the bump in the binary source model will be significantly different from the true source producing the planetary perturbation (or vice versa). Because $\theta_* \propto \sqrt{\mathrm{flux}}$, if the inferred source is much fainter than the true source, it should have a much smaller $\theta_*$. Smaller values for $\theta_*$ imply a value of $\mu_{\rm rel}$ that is smaller by the same factor. 

\begin{figure*}
	\includegraphics[width=\textwidth]{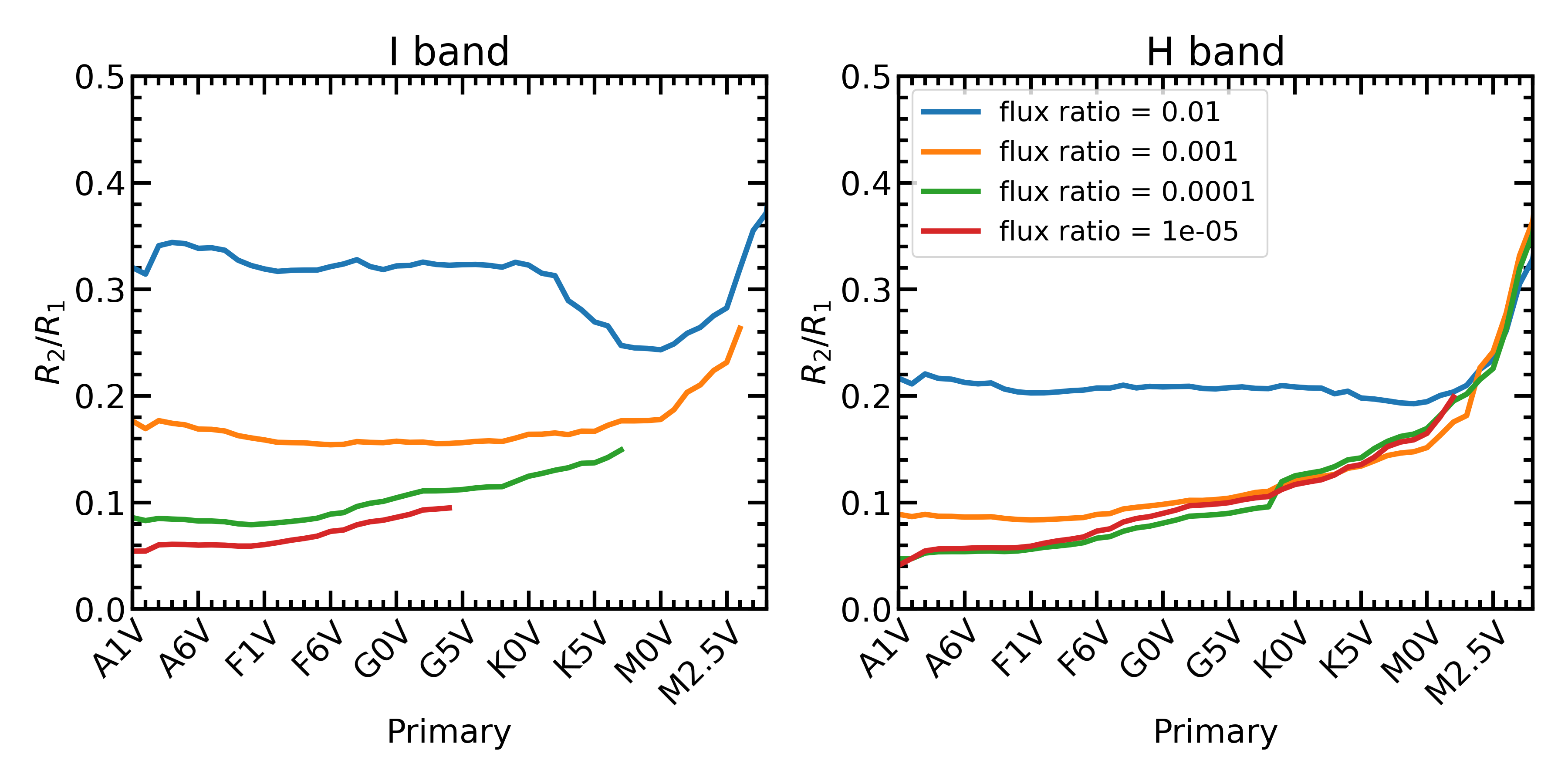}
	\caption{Radius of the secondary relative to the radius of the primary for binary sources with various flux ratios (see legend) in $I$ (left) and $H$ bands \citep{PecautMamajek13}. \label{fig:radius_ratios}}
\end{figure*}

Figure \ref{fig:radius_ratios} shows the ratio of the radius of the implied secondary to the radius of the primary as a function of spectral type for a few specific flux ratios \citet{PecautMamajek13}. These calculations assume both stars are on the main sequence, which will be true for Roman. We also assume the properties of the primary source in the binary source model is the same as the true source from the planetary case because the main stellar event is nearly the same in either case.
The figure suggests the implied secondary from a binary source model will usually be a factor of 10 to 20 smaller than the source creating a planetary event. 

Typical values of $\mu_{\rm rel}$ are 1 to 10 mas~yr$^{-1}$. Values of $\mu_{\rm rel} \lesssim 1~\mathrm{mas~yr}^{-1}$ are extremely unlikely \citep{Gould22_MASADA,Jung23AF8}, while values of $\mu_{\rm rel}$ much larger than $10~\mathrm{mas~yr}^{-1}$ would imply a star that is unbound from the Galaxy. So even if the light curve models are degenerate, the binary source model might be excluded or disfavored because the implied $\mu_{\rm rel}$ is too small.

  We can assume the properties of the primary source to be the same as the true source from the planetary case because the main stellar event is nearly the same in either case. In the case of RGBS, the primary sources will be constrained to be main sequence stars.  Figure \ref{fig:radius_ratios} shows the ratio of the radius of the implied secondary to the radius of the primary as a function of spectral type for a few specific flux ratios \citet{PecautMamajek13}. The implied secondary will usually be a factor of 10 to 20 smaller than the primary. 

\begin{figure*}
	\includegraphics[width=\textwidth]{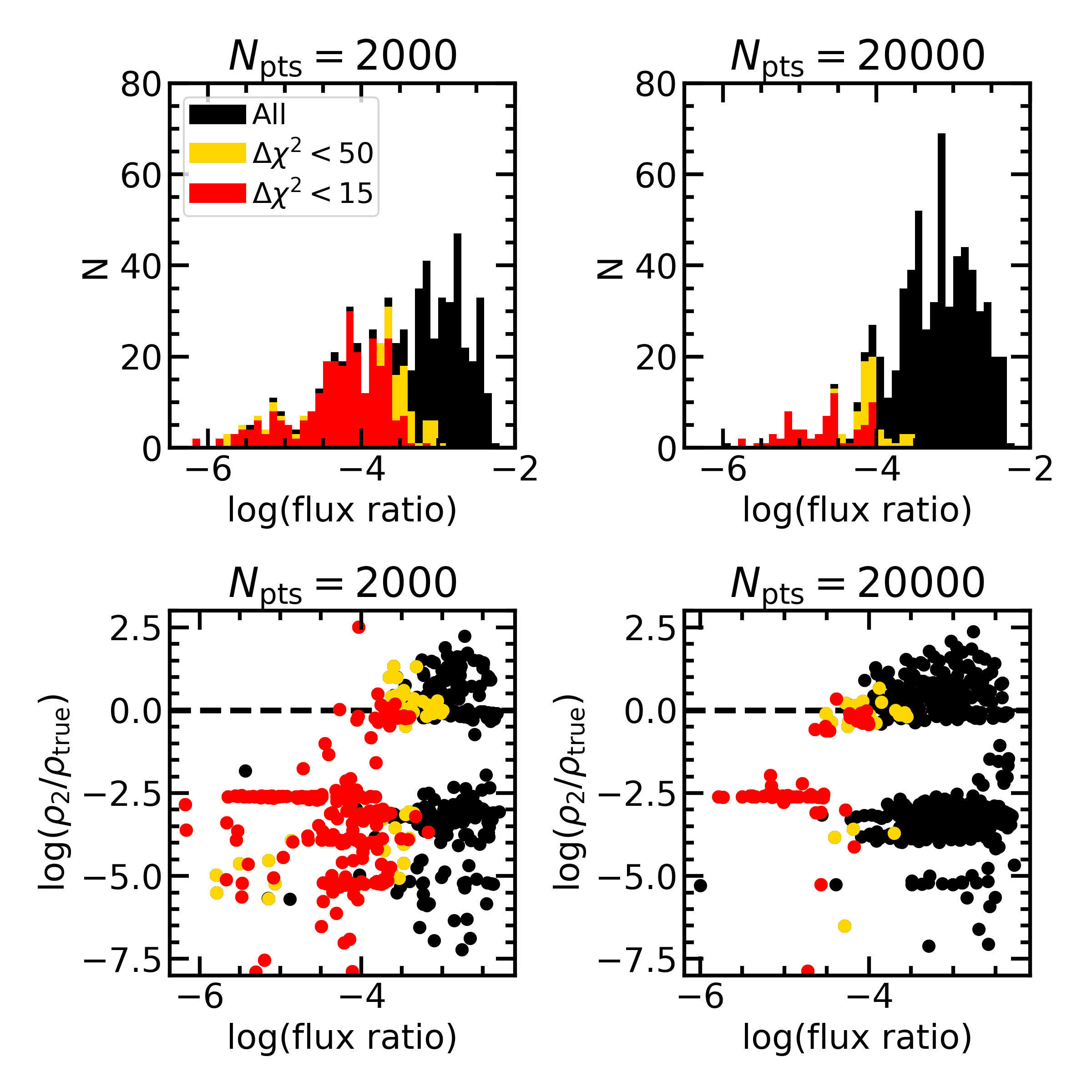}
	\caption{{\em Top:} Distributions of the logarithms of the source flux ratio for the best-fitting binary source models. {\em Bottom:} logarithms of the source flux ratio compared to the best-fitting source size for the second source compared to the true value from the planetary model. Cases where the binary source light curve model is degenerate with the planet at  $\Delta\chi^2 = 15, 50$ are shown in red, gold. The left panels show the results for $N_{\rm pts}=2000$ while the right panels show our fiducial case with $N_{\rm pts}=20000$. \label{fig:flux_ratios}}
\end{figure*}

  To test this hypothesis, we compare the fitted flux ratio and the ratio of the fitted value of $\rho_2$ to the true value, $\rho_{\rm true}$, in Figure \ref{fig:flux_ratios}, . The points are color-coded by the $\Delta\chi^2$ of the binary source solution. For simplicity, we only consider the $N=2000$ and $N=20000$ simulations for this Figure. When the degeneracy is unresolved at the $\Delta\chi^2 = 15$ level, the secondary is always $10^{-3}$ times fainter that the primary and frequently $10^{-4}$ times fainter, consistent with theoretical predictions. 

  However, these degenerate cases usually correspond to non-caustic-crossing perturbations. The bottom panels of Figure \ref{fig:flux_ratios} shows that the best-fitting source size for the secondary in the binary source models is frequently a factor of 100 or more smaller than the source used to generate the simulated planetary data. We infer that this indicates the size of the binary source is not measured. Without a measurement of $\rho_2$, there is only a lower-limit on $\mu_{\rm rel}$. 

  In Table \ref{tab:mu_rel}, we quantify the number of light curves with the a degenerate binary source model as a function of $N_{\rm pts}$ and $\Delta\chi^2$ as well as the number of binary source models with a measured value of $\rho_2$. Based on the divisions between clusters of points in Figure \ref{fig:flux_ratios}, we assume that $\rho_2$ is measured when $\log(\rho_2 / \rho_{\rm true}) > -0.5$. This will tend to over-estimate the number of solutions with a measured $\rho_2$, and so form an upper limit. Finally, we use the \citet{PecautMamajek13} tables to calculate the ratio, $f_\mu$, between the value of $\mu_{\rm rel}$ that would be derived from the source flux, $\rho$, and $t_{\rm E}$ from the binary source model and the true value. For these calculations, we consider the case of either a G5 or M0 primary and observations in either the $I$ or $H$-band. 

  Table \ref{tab:mu_rel} shows that very few planetary events with degenerate binary source models will have measured values for $\rho_2$ because most degenerate models come from the non-caustic crossing \textit{traj 3} simulations. Our calculations of the ratio of $\mu_{\rm rel}$ show that even fewer are likely to have extreme values of $\mu_{\rm rel}$. As can been seen in the lower-right panel of Figure \ref{fig:flux_ratios}, degenerate events tend to have values of $\rho_2$ that are smaller than the true value, partially compensating for flux effect discussed above. So, while there are some degenerate cases that may be resolved based on the improbability from the derived $\mu_{\rm rel}$, such cases will be relatively rare.

\begin{table*}
\caption{Light curves with degenerate 1L2S models\label{tab:mu_rel}}
\begin{tabular}{|r|r|rrr|rrrr|rrrr|}
\hline
\hline
\multicolumn{13}{|c|}{1\% photometry}\\
\hline
\multicolumn{1}{|c|}{} & \multicolumn{1}{c|}{} & \multicolumn{3}{c|}{}& \multicolumn{4}{c|}{$|\log(f_\mu)|>0.7$}& \multicolumn{4}{c|}{$|\log(f_\mu)|>1.0$}\\
\multicolumn{1}{|c|}{} & \multicolumn{1}{c|}{$\rho$}  &\multicolumn{2}{c}{} &  \multicolumn{1}{c|}{AND $\rho$} & \multicolumn{2}{c}{G5}& \multicolumn{2}{c|}{M0}& \multicolumn{2}{c}{G5}& \multicolumn{2}{c|}{M0}\\
$N_{\rm pts}$ & meas. & $\Delta\chi^2$ & $N < \Delta\chi^2$ & meas.& I & H & I & H & I & H & I & H \\
\hline\hline
  80k & 268 &    50 &     68 &   22 &   9 &   9 &   4 &   4 &   3 &   3 &   1 &   1\\
\multicolumn{1}{|c|}{} & \multicolumn{1}{c|}{}  &    15 &     41 &    3 &   0 &   0 &   0 &   0 &   0 &   0 &   0 &   0\\
\hline
  40k & 281 &    50 &     95 &   45 &  26 &  27 &  10 &   8 &   6 &   7 &   1 &   1\\
\multicolumn{1}{|c|}{} & \multicolumn{1}{c|}{}  &    15 &     57 &   11 &   1 &   1 &   1 &   1 &   0 &   0 &   0 &   0\\
\hline
  20k & 298 &    50 &    114 &   59 &  40 &  43 &  22 &  16 &   9 &  12 &   5 &   5\\
\multicolumn{1}{|c|}{} & \multicolumn{1}{c|}{}  &    15 &     70 &   19 &   8 &  10 &   3 &   2 &   1 &   2 &   1 &   1\\
\hline
   5k & 353 &    50 &    196 &   89 &  67 &  78 &  30 &  26 &  15 &  23 &   8 &   6\\
\multicolumn{1}{|c|}{} & \multicolumn{1}{c|}{}  &    15 &    126 &   42 &  30 &  32 &  11 &   8 &   4 &   8 &   3 &   2\\
\hline
   2k & 197 &    50 &    351 &   73 &  58 &  70 &  34 &  37 &  20 &  26 &  14 &  14\\
\multicolumn{1}{|c|}{} & \multicolumn{1}{c|}{}  &    15 &    286 &   38 &  26 &  36 &   9 &   9 &   5 &   7 &   3 &   3\\
\hline
   1k & 140 &    50 &    486 &   72 &  38 &  66 &  32 &  34 &  23 &  28 &  21 &  21\\
\multicolumn{1}{|c|}{} & \multicolumn{1}{c|}{}  &    15 &    395 &   30 &  15 &  25 &  11 &  10 &   8 &   8 &   6 &   6\\
\hline
\hline
\multicolumn{13}{|c|}{5\% photometry}\\
\hline
\multicolumn{1}{|c|}{} & \multicolumn{1}{c|}{} & \multicolumn{3}{c|}{}& \multicolumn{4}{c|}{$|\log(f_\mu)|>0.7$}& \multicolumn{4}{c|}{$|\log(f_\mu)|>1.0$}\\
\multicolumn{1}{|c|}{} & \multicolumn{1}{c|}{$\rho$}  &\multicolumn{2}{c}{} &  \multicolumn{1}{c|}{AND $\rho$} & \multicolumn{2}{c}{G5}& \multicolumn{2}{c|}{M0}& \multicolumn{2}{c}{G5}& \multicolumn{2}{c|}{M0}\\
$N_{\rm pts}$ & meas. & $\Delta\chi^2$ & $N < \Delta\chi^2$ & meas.& I & H & I & H & I & H & I & H \\
\hline\hline
  80k & 268 &    50 &    190 &  111 &  79 &  91 &  38 &  37 &  23 &  30 &   9 &  10\\
\multicolumn{1}{|c|}{} & \multicolumn{1}{c|}{}  &    15 &    134 &   75 &  50 &  56 &  22 &  21 &  14 &  16 &   3 &   3\\
\hline
  40k & 281 &    50 &    229 &  144 & 108 & 122 &  64 &  63 &  39 &  51 &  24 &  23\\
\multicolumn{1}{|c|}{} & \multicolumn{1}{c|}{}  &    15 &    169 &  106 &  73 &  85 &  34 &  31 &  21 &  28 &   7 &   6\\
\hline
  20k & 298 &    50 &    295 &  184 & 142 & 166 &  88 &  92 &  56 &  68 &  42 &  44\\
\multicolumn{1}{|c|}{} & \multicolumn{1}{c|}{}  &    15 &    198 &  123 &  90 & 105 &  52 &  47 &  28 &  37 &  16 &  16\\
\hline
   5k & 353 &    50 &    470 &  257 & 208 & 241 & 151 & 158 & 116 & 134 & 102 & 103\\
\multicolumn{1}{|c|}{} & \multicolumn{1}{c|}{}  &    15 &    357 &  194 & 158 & 181 & 101 & 107 &  68 &  85 &  56 &  56\\
\hline
   2k & 197 &    50 &    607 &  173 & 125 & 168 &  93 & 109 &  71 &  82 &  62 &  66\\
\multicolumn{1}{|c|}{} & \multicolumn{1}{c|}{}  &    15 &    510 &  146 & 111 & 143 &  79 &  93 &  57 &  68 &  48 &  52\\
\hline
   1k & 140 &    50 &    665 &  138 &  72 & 108 &  66 &  69 &  55 &  62 &  52 &  53\\
\multicolumn{1}{|c|}{} & \multicolumn{1}{c|}{}  &    15 &    611 &  120 &  65 & 101 &  59 &  62 &  48 &  55 &  45 &  46\\
\hline
\hline
\end{tabular}\tablecomments{There are 675 total simulated light curves for each value of $N_{\rm pts}$. Light curves for which the best-fitting 1L2S model has $\log(\rho_2/ \rho_{\rm True}) > -0.5$ are considered to have $\rho$ measured. We consider two possible thresholds for the 1L2S degeneracy: $\Delta\chi^2 = (15, 50)$. The number of light curves for which the degeneracy exists is $N < \Delta\chi^2$. The intersection of that set and those with measured $\rho$ is given in the column labeled ``AND $\rho$ meas." The final columns give the number of degenerate 1L2S models that result in a derived value of $\mu_{\rm rel}$ that is more than a factor of 5 or 10 ($|\log(f_{\mu})| > $ 0.7 or 1.0, respectively) different from the true value. The simulations were conducted using 1\% photometry (top section). The 5\% photometry section (bottom) shows the results after dividing the $\Delta\chi^2$ by 25.}
\end{table*}

\section{Scaling the Results}

Our calculations were carried out for $t_{\rm E} = 20~\mathrm{day}$ and assuming 1\% photometric precision. These results can be scaled and interpolated to assess the impact of degeneracy under different values of observing cadence, $t_E$, or photometric precision. 

\begin{figure*}
	\centering	
	\includegraphics[width=\textwidth]{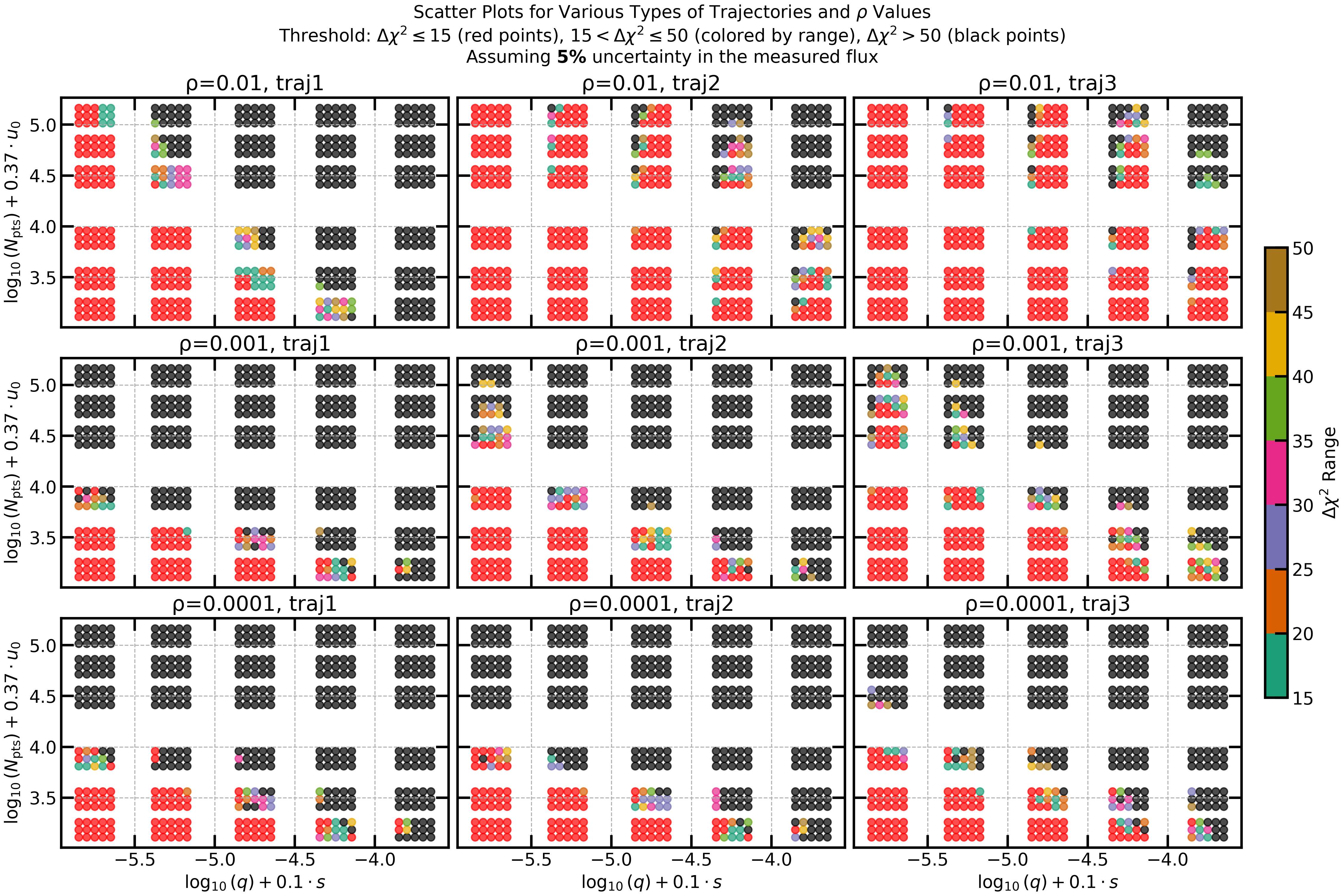}
	\caption{This figure is identical to Figure~\ref{fig1t}, except that here the photometric precision is set to 5\% by dividing $\Delta\chi^2$ by 25.}
	\label{fig2t}
\end{figure*}

For instance, to modify the photometric precision, one simply needs to adjust $\Delta\chi^2$. As an example, dividing $\Delta\chi^2$ by 25 scales the results to a photometric precision of 5\% instead of 1\%. Applying this adjustment to Figure~\ref{fig1t}, which was originally generated for a photometric precision of 1\%, yields the results for 5\% photometric precision as presented in Figure~\ref{fig2t} and the bottom section of Table~\ref{tab:mu_rel}. 

As expected, more planets have degenerate binary source solutions with worse photometric precision. However, even with this factor, if $N_{\rm pts}=20,000$, the degeneracy is mostly resolved except for the smallest mass ratios or largest sources. In addition, some of the degeneracies may be resolved by measuring $\mu_{\rm rel}$.

We provide our complete results as Supplementary Data to enable the reader to scale the results as needed for other investigations.

\section{Conclusions}

  We investigated the binary source degeneracy \citep{gaudi1998distinguishing} for simulated observations of wide-orbit planets assuming 1\% photometric precision. Our analysis demonstrates that the degeneracy is rarely a problem if the observing cadence is $\ge (0.001t_{\rm E})^{-1}$ $\sim (15~\mathrm{min})^{-1}$ for $t_{\rm E} = 20~$ day. The exception is events with large sources. For $\rho = 0.01$, binary source models are degenerate for mass ratios $q \lesssim 10^{-4}$ unless the source trajectory crosses directly through the planetary caustic. Fortunately, most source stars in the Roman Galactic Exoplanet Survey are likely to be dwarfs with angular source radii $\sim 0.5~\mu\mathrm{as}$. For typical values of $\theta_{\rm E}\sim 0.5~\mathrm{mas}$, this means that $\rho \sim 0.001$, so this degeneracy is not likely to be an issue for Roman microlensing events with 1\% photometry.

  The binary source degeneracy is potentially a problem for ground-based microlensing surveys. Ground-based microlensing surveys are typically conducted with 1m- to 2m-class telescopes \citep{Udalski03,Bond01,Kim16_KMTNet} that achieve photometric precisions of 1\% only for microlensing events with bright (i.e., giant) sources or high-magnification events. Because giant sources tend to have $\rho \gtrsim 0.01$, our investigation shows that the binary source degeneracy creates challenges for detecting planets smaller than $\log q= -4.5$ unless the cadence of observations is high ($\ge (15~\mathrm{min})^{-1}$); the source must also cross the planetary caustic.
So, although the large size of giant source stars is an advantage for increasing the probability of a planetary caustic interaction, the binary source degeneracy could create ambiguity in interpreting the detected signals.

\section*{Acknowledgements}

J.C.Y. acknowledges support from U.S. NASA Grant No. 22-RMAN22-0078.

  This research has made use of the NASA Exoplanet Archive,  
which is operated by the California Institute of Technology,  
under contract with the National Aeronautics and Space Administration  
under the Exoplanet Exploration Program.  

  \textbf{Facility:} Exoplanet Archive



\bibliographystyle{aasjournal}

\end{document}